\documentstyle[prb,aps,epsfig,multicol]{revtex}
\begin{document}
\draft
\title{Crystallization of a polymer on a surface}
\author{Jonathan P.~K.~Doye and Daan Frenkel}
\address{FOM Institute for Atomic and Molecular Physics, Kruislaan 407,\\ 
1098 SJ Amsterdam, The Netherlands}
\date{\today}
\maketitle
\begin{abstract}
We have studied the structure and free energy landscape of a semi-flexible lattice-polymer
in the presence of a surface. 
At low temperatures coexistence of two-dimensional integer-folded crystals is observed. 
As the temperature is increased there is a transition from these 
crystalline configurations to a disordered coil adsorbed onto the surface.
The polymer then gradually develops three-dimensional character at higher temperatures. 
We compute the free energy as a function of increasing crystallinity and 
compare with the free energy profiles assumed by the Lauritizen-Hoffman surface nucleation theory
of polymer crystallization.
Our free energy profiles exhibit a `sawtooth' structure associated with the successive 
formation of chain folds.
However, in the early stages of crystallization our profiles significantly deviate 
from those assumed by surface nucleation theory because the initial nucleus is not 
a single stem but two incomplete stems connected by a fold. 
This finding has significant implications for the theoretical description of polymer crystallization.
\end{abstract}
\begin{multicols}{2}
\section{Introduction}
On crystallization from solution or the melt simple polymers typically 
form lamellar crystals. As the backbone of the polymer chain is oriented 
perpendicular to the plane of the lamellar, and yet the thickness of the crystals are
smaller than the length of the chain, a single polymer must
traverse the crystal many times folding back on itself at each surface.\cite{Keller57a}.
Furthermore the thickness of the crystals have a well-characterized 
dependence on the degree of supercooling. The thickness is always slightly
larger than the minimum thickness for which a lamellar crystal is thermodynamically stable.

However, although these simple facts were discovered about 40 years ago, there
is still no consensus on their theoretical explanation.\cite{theorynoteb} 
Moreover, two of the more dominant approaches---Lauritzen-Hoffman surface nucleation 
theory\cite{Lauritzen60,Hoffman76a,Hoffman97}
and the entropic barrier model\cite{Sadler84a,Sadler86a,Sadler88a,Spinner95}---appear irreconcilable.
One of the difficulties is that the theories often have to make assumptions about
the microscopic mechanisms of crystallization. 
As it is difficult to probe these processes directly in an experiment, 
the main test for the theories has been the comparison of their predictions
with macroscopic properties, such as the crystal thickness, growth rate and shape. 
But these tests have not proved discriminating enough; 
the two main theories (perhaps with the help of various refinements or `suitable' choices of parameter)
have both been able to provide an adequate description of many of these properties.

Therefore, (atomistic) simulation can potentially play an important role 
in this debate by providing insight into the microscopic processes involved 
in polymer crystallization,
and thus help in the critical assessment and refinement of the current theories 
and perhaps in the development of new theories.
However, there have been few simulation studies\cite{Yamamoto97,Chen98} which have attempted 
to investigate the basic process in solution crystallization, the adsorption {\it and}
crystallization of a polymer on the growth surface. 
(There are many studies which just consider
adsorption.\cite{Eisenriegler93,Eisenriegler82,Meirovitch88,Chang92,Foster92,Debell93,Lai94,Milchev93b})
Although this is a non-equilibrium process, as a first step it is useful to understand
the equilibrium behaviour of a single polymer in the presence of a surface.
In this paper, we first study the basic thermodynamic 
properties of a simple lattice model of such a system (Section \ref{sect:thermo}).
This will also add to the increasing knowledge of the rich phase behaviour of single homopolymer chains;
in particular, we draw out the similarities and differences from the behaviour of an isolated semi-flexible
polymer, a case that has received much more 
attention.\cite{Kolinski86a,Kuznetsov96a,Zhou96a,Doniach96a,Bastolla97a,Fujiwara97a,Doye98a,Noguchi97}
Secondly, we examine in detail the free energy profile for the crystallization pathway
suggested by surface nucleation theory (Section \ref{sect:profiles}).

\section{Methods}
\label{ref:meth}
\subsection{Polymer Model}
\label{sect:pmodel}
In our model the polymer is represented by an $N$-unit self-avoiding walk 
on a simple cubic lattice. 
There is an attractive energy, -$\epsilon$, between non-bonded polymer units on 
adjacent lattice sites and between polymer units and the surface, 
and an energetic penalty, $\epsilon_g$, for kinks in the chain. 
The total energy is given by
\begin{equation}
E=-(n_{pp}+n_{ps})\epsilon+n_g\epsilon_{g}
\label{eq:energy}
\end{equation}
where $n_{pp}$ is the number of polymer-polymer contacts, $n_{ps}$ is the 
number of polymer-surface contacts
and $n_g$ is the number of kinks or `gauche bonds' in the chain.
$\epsilon$ can be considered to be an effective interaction representing 
the combined effects of polymer-polymer, polymer-solvent and solvent-solvent interactions,
and so our model is a simplified representation of a semi-flexible polymer at 
the interface between a polymer crystal and solution.
The behaviour of the polymer is controlled by the ratio $kT/\epsilon$; 
large values can be considered as either high temperature or good solvent conditions, 
and low values as low temperature or bad solvent conditions.
The parameter $\epsilon_g$ defines the stiffness of the chain. The polymer
chain is fully flexible at $\epsilon_g$=0 and becomes stiffer as $\epsilon_g$ increases.
In this study we only consider $\epsilon_g\geq 0$.

This polymer model has been recently used to study
the effects of stiffness on the phase behaviour of isolated homopolymers by 
theory\cite{Doniach96a} and simulation\cite{Bastolla97a,Doye98a}, and
also in kinetic Monte Carlo simulations of the growth of
polymer crystals.\cite{Doye98b}

The global potential energy minimum at a particular (positive) $\epsilon_{g}$ is 
determined by a balance between maximizing $n_{pp}$ and $n_{ps}$,
and minimizing $n_{g}$; it is a folded structure that lies flat on the surface. 
If the polymer is able to form a structure that is a 2-dimensional rectangle with
dimensions $a\times b$ ($N=ab$), where
$a\le b$, then 
\begin{eqnarray}
n_{ps}&=&N \nonumber \\
\label{eq:ng}
n_{pp}&=&N-a-b+1 \\
n_{g}^{\rm min}&=&2b-2 \nonumber
\end{eqnarray}
The structures that correspond to $n_g=n_g^{\rm min}$ have the polymer chain
folded back and forth along the longer dimension of the rectangle.
By minimizing the resulting expression for the energy one
finds that the lowest energy polymer configuration should have 
\begin{equation}
{b\over a}=1+{2\epsilon_{g}\over\epsilon}.
\label{eq:aspect}
\end{equation}
Therefore, at $\epsilon_{g}=0$ the ideal shape is a square and for 
positive $\epsilon_{g}$ a rectangle extended in one direction, 
the aspect ratio of which increases as the chain becomes stiffer.
However, at most sizes and values of $\epsilon_{g}$ it is not possible 
to form a rectangle with the optimal aspect ratio. 
Nevertheless, it is easy to find the global minimum just 
by considering the structures which most closely approximate this optimal shape.

\subsection{Simulation Techniques}

To simulate our system we use configurational-bias
Monte Carlo\cite{Siepmann92a} including moves in which a mid-section 
of the chain is regrown.\cite{Dijkstra94a} 
We also make occasional bond-flipping moves which, 
although they do not change the shape of the volume occupied by the polymer, 
change the path of the polymer through that volume.\cite{Doye98a,Ramakrishnan97a,Deutsch97a}
These moves speed up equilibration in the dense phases.
During the simulation we always constrain the polymer to have one unit 
in contact with the surface. 
Thermodynamic properties, such as the heat capacity, were calculated from the energy 
distributions of each run using the multi-histogram method.\cite{Ferrenberg89a,Poteau94a}

In order to monitor the state of the polymer we devised two order parameters.
The first, $Q_{2D}$, probes the orientational order within the 2-dimensional polymer
adsorbed onto the surface.
\begin{equation}
Q_{2D}=\sqrt{2\sum_{\alpha=x,y} \left({n_\alpha\over n_x+n_y}-{1\over 2}\right)^2},
\end{equation}
where $x$ and $y$ are in the plane of the surface and
$n_\alpha$ is the number of bonds in the direction $\alpha$. 
$Q_{2D}$ has a value of 1 if all the bonds are in the same direction, i.e.\ the polymer 
has a linear configuration, and a value of 0 if the bonds are oriented isotropically in the plane. 
The second order parameter, $Q_{3D}$, probes the dimensionality of the polymer.
\begin{equation}
Q_{3D}=\sqrt{6 \left[2\left({n_x+n_y\over 2(N-1)}-{1\over 3}\right)^2+
                       \left({n_z\over (N-1)}-{1\over 3}\right)^2 \right]}.
\end{equation}
$Q_{3D}$ has a value of 1 if all the polymer units are in contact with the surface and 
a value of 0 if the bonds are oriented isotropically in space.

In section \ref{sect:profiles} we compute the free energy along pathways characterized 
by an order parameter, $N_{\rm xtal}$, which measures the degree of crystallinity. 
This Landau free energy is simply related to the canonical probability distribution for the 
order parameter: $A_L(N_{\rm xtal})=A-kT \log p_{can} (N_{\rm xtal})$, where $A$ is the 
Helmholz free energy. 
However, all relevant regions of this distribution are not significantly sampled in 
the canonical ensemble, and so we use umbrella sampling\cite{Torrie77} to calculate the
free energy accurately over the whole range of the order parameter. 
This is achieved by multiplying the Boltzmann factor 
by the exponential of a biasing distribution, $W(N_{\rm xtal})$; 
i.e.\ the simulation samples configurations with a probability proportional to
$\exp(-\beta E + W(N_{\rm xtal}))$. 
The canonical probability distribution is then obtained from 
the probability distribution from the biased run, $p_{\rm multi}(N_{\rm xtal})$, by 
$p_{can}(N_{\rm xtal})=p_{\rm multi}(N_{\rm xtal}) \exp(-W(N_{\rm xtal}))$.
We wish to choose $W$ such that $p_{\rm multi}(N_{\rm xtal})$ is approximately 
constant over the whole range of $N_{\rm xtal}$ 
(the so-called multicanonical approach\cite{Berg91,Berg92}).
However, this only occurs when $W(N_{\rm xtal}) \approx A_L(N_{\rm xtal})/kT$ 
and so we have to construct $W$ iteratively from the results of a number of 
short preliminary simulations.\cite{Berg96}

\section{Thermodynamics}
\label{sect:thermo}

In the presentation of our results we concentrate on one example, 
a 200-unit polymer with $\epsilon_g=4\epsilon$. Results for other positive
values of $\epsilon_g$ show the same basic behaviour. 
We also mainly dwell on those aspects of the thermodynamics which 
are relevant to polymer crystallization
or 
differ markedly from the behaviour of the isolated 
homopolymer.\cite{Doniach96a,Bastolla97a,Doye98a}

\end{multicols}
\begin{figure}
\begin{center}
\vglue -1.2cm
\epsfig{figure=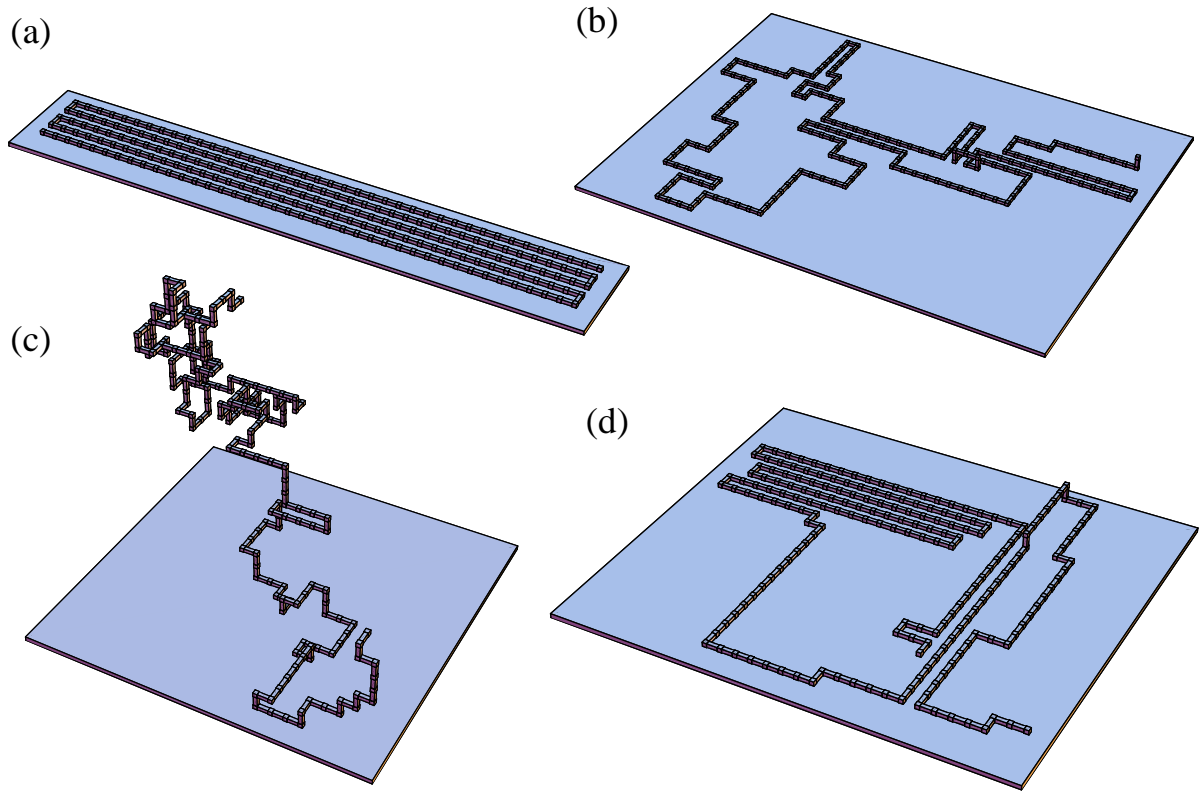,width=17cm}
\vglue -1.4cm
\begin{minipage}{18cm}
\caption{\label{fig:piccies} 
Some representative configurations for a 200-unit polymer 
in the presence of a surface.
(a) Crystalline configuration with 5 stems of length 40 units;
it is one of the lowest energy configurations for $\epsilon_g=4\epsilon$. 
$Q_{2D}$=0.960 , $R_g^2$=135.3.
(b) Disordered two-dimensional coil on the surface produced in a run at 
$T=2.5\epsilon k^{-1}$. $Q_{2D}$=0.235 , $R_g^2$=141.6.
(c) Three-dimensional coil produced in a run at 
$T=5.0\epsilon k^{-1}$. $Q_{3D}$=0.010, $R_g^2$=99.1.
(d) Configuration from a simulation at $T=2.375 \epsilon k^{-1}$ in which half the 
configuration is crystalline and half is disordered. 
}
\end{minipage}
\end{center}
\end{figure}
\begin{multicols}{2}

\begin{figure}
\begin{center}
\epsfig{figure=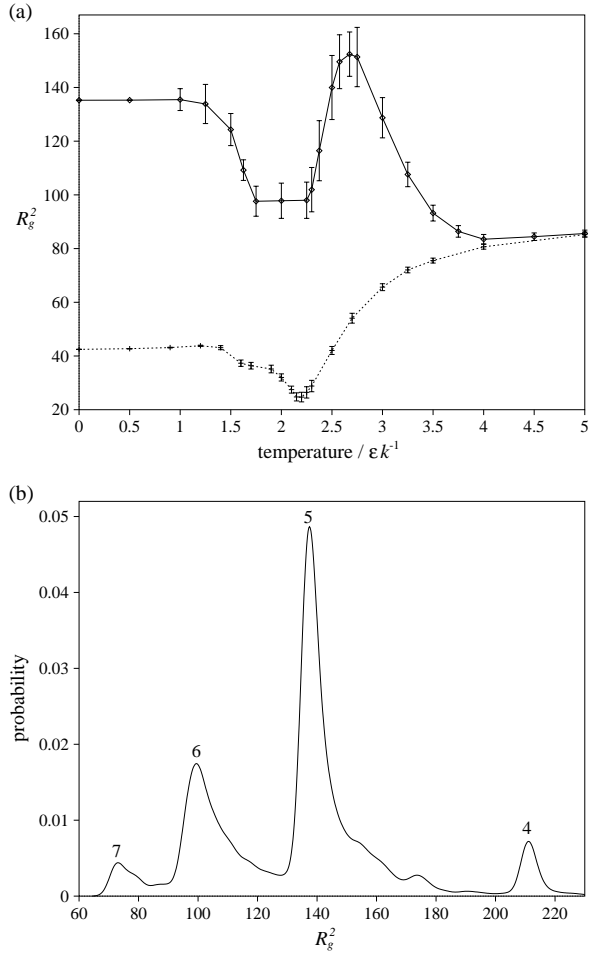,width=8.2cm}
\vglue0.1cm
\begin{minipage}{8.5cm}
\caption{\label{fig:rg} (a) $R_g^2$ as a function of temperature for a 200-unit 
polymer at $\epsilon_g=4\epsilon$. 
For comparison the results for an isolated polymer (dashed line) have also been included.
(b) Probability distribution for $R_g^2$ at $T=1.25\epsilon k^{-1}$.
The peaks are labelled by the numbers of stems in the corresponding configurations.}
\end{minipage}
\end{center}
\end{figure}

At zero temperature the global potential energy minimum has the lowest free energy.
For $\epsilon_g=4\epsilon$ the optimal aspect ratio of a crystal is 9 (Equation (\ref{eq:aspect})).
For $N$=200 the crystal that most closely approximates this shape has five stems 
(a stem is a straight section of the polymer) which are 40 units long; it is the global minimum. 
The example shown in Figure \ref{fig:piccies}a
has a tight fold at the end of each stem, a situation which is often referred to as adjacent 
re-entry, but there are a number of degenerate configurations with non-adjacent re-entry. 
The differences in energy between crystalline configurations 
are only small and so at low temperatures more than one type of crystalline 
configuration is observed in the simulation; 
they can be differentiated by their radius of gyration. 
Figure \ref{fig:rg}b shows the presence of four coexisting crystallites which have four,
five, six or seven complete stems. This preference for crystalline configuration 
with complete stems is energetic in origin---it maximizes the number
of polymer-polymer contacts for each fold. 
A similar trend is observed in long monodisperse alkanes; 
the crystals have preferred thicknesses which correspond to an integer number 
of complete stems.\cite{Ungar85,Organ96}

As the temperature is increased configurations with more disorder in the stem length and 
a shorter average stem length are entropically more favoured. 
This causes the peaks in the probability distribution of the radius of gyration to broaden 
and the radius of gyration to decrease in the range $T=1.0$--$1.75\epsilon k^{-1}$ 
(Figure \ref{fig:rg}a).

\begin{figure}
\begin{center}
\epsfig{figure=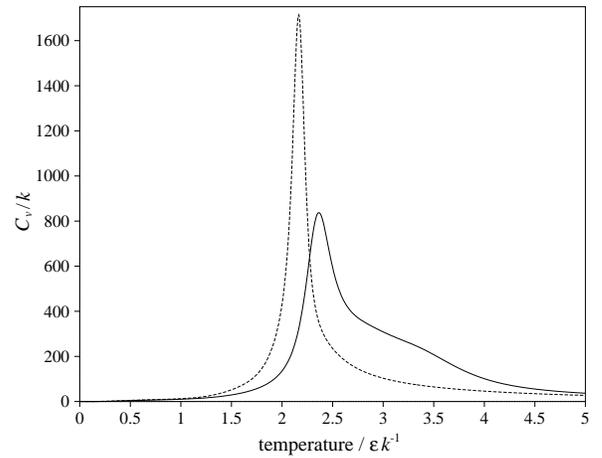,width=8.2cm}
\vglue0.1cm
\begin{minipage}{8.5cm}
\caption{\label{fig:cv} $C_v$ as a function of temperature for a 200-unit 
polymer at $\epsilon_g=4\epsilon$.
For comparison the results for an isolated polymer (dashed line) have also been included.
}
\end{minipage}
\end{center}
\end{figure}

\begin{figure}
\begin{center}
\epsfig{figure=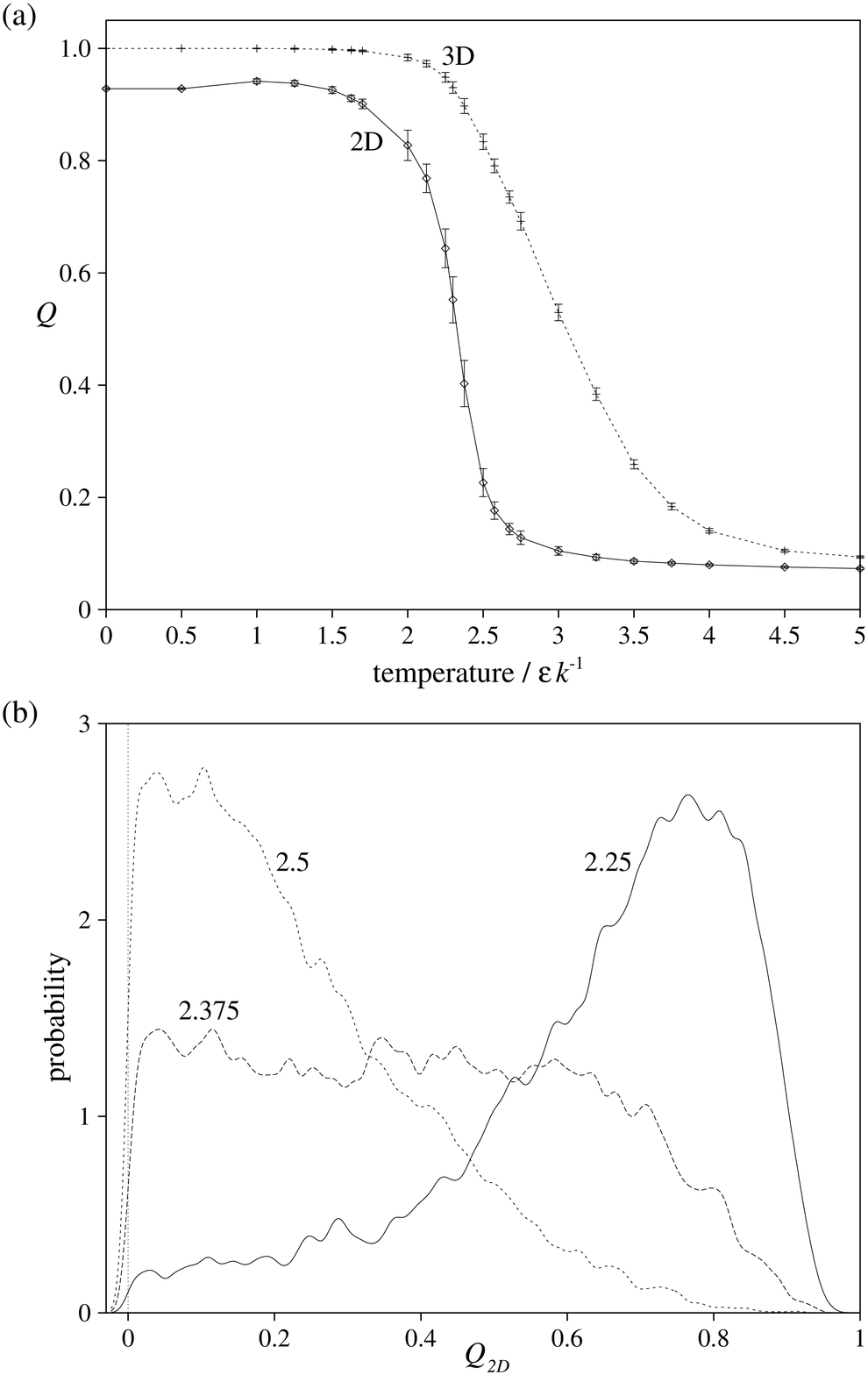,width=8.2cm}
\vglue0.1cm
\begin{minipage}{8.5cm}
\caption{\label{fig:op} (a) $Q_{2D}$ and $Q_{3D}$ as a function of temperature for a 200-unit 
polymer at $\epsilon_g=4\epsilon$.
(b) Probability distributions for $Q_{2D}$ at three different temperatures around the
melting temperature, as labelled.
}
\end{minipage}
\end{center}
\end{figure}

As with the isolated homopolymers there must come a point when disordered configurations
become more favoured than crystalline configurations. 
This `melting' transition is signalled by a heat capacity peak (Figure \ref{fig:cv})
and by a loss of orientational order (Figure \ref{fig:op}a).
Although $Q_{2D}$ drops down to a value of 0.1 at the transition, 
$Q_{3D}$ remains close to one. 
This shows that on melting the polymer adopts a two-dimensional configuration 
on the surface with no orientational order in the plane (Figure \ref{fig:piccies}b).
Furthermore, although this order-disorder transition is a finite size analogue of a bulk first-order
phase transition, no bimodality is seen in the probability distribution of 
$Q_{2D}$ (Figure \ref{fig:op}b), 
i.e. there is no free energy barrier between states with high and low $Q_{2D}$.
Near to the transition temperature, $T_m$, the $Q_{2D}$ probability distribution is very broad and flat,
perhaps because of the contribution of partly crystalline, partly disordered configurations 
(such as the one depicted in Figure \ref{fig:piccies}d) with intermediate values of $Q_{2D}$.

The order-disorder transition also causes the radius of gyration to increase 
first as the polymer passes to the disordered state and 
then as this coil begins to expand with temperature. 
However, this rise is checked when the polymer begins to develop a three-dimensional character 
(Figure \ref{fig:piccies}c).
The change in dimensionality occurs gradually and is signalled by a decrease in $Q_{3D}$
and also a decrease in the radius of gyration to a value which is comparable to that for the
an isolated chain of the same size (Figure \ref{fig:rg}a).
The associated loss in polymer-surface contacts is the cause of the high temperature
shoulder in the heat capacity. 
Such desorbtion transitions are well-understood\cite{Eisenriegler93} and 
we do not dwell on its features here.

In Figures \ref{fig:rg} and \ref{fig:cv} we have included results for an isolated polymer for comparison.
Until the transition to a three-dimensional configuration, the $R_g^2$ curves for the two cases
have a similar form (except that the values are larger for the 
2D polymer on the surface) due to the similar nature of the transitions in the two cases.\cite{Doye98b} 
The heat capacity peak for the isolated polymer is sharper, suggesting that the melting
transition has a stronger first-order-like character. 
Similarly, it is easier to find an order parameter for the isolated polymer
that shows bimodality (i.e.\ the presence of free energy barrier) in the transition region.
This difference is probably because the energy difference between ordered and disordered states
is less in the presence of the surface; the energy of the two-dimensional coil is reduced relative
to the three-dimensional coil due to the interaction energy with the surface.

\begin{figure}
\begin{center}
\epsfig{figure=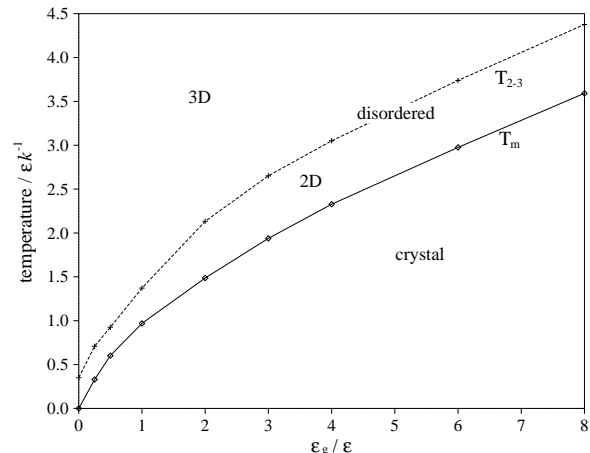,width=8.2cm}
\vglue0.1cm
\begin{minipage}{8.5cm}
\caption{\label{fig:phase}Phase diagram of a 200-unit polymer on a surface.
The phase diagram is  divided into regions by the values of
the melting point, $T_m$, and by the transition from two- to
three-dimensional configurations, $T_{2-3}$. 
These two transition temperatures are defined by
$Q_{2D}(T_m)=0.5$ and $Q_{3D}(T_{2-3})=0.5$.
No attempt has been made to differentiate between coils and globules in the
disordered region of the phase diagram.}
\end{minipage}
\end{center}
\end{figure}

The basic behaviour described above generally holds for any positive $\epsilon_g$,
as be can seen from the phase diagram in Figure \ref{fig:phase}.
In particular the order-disorder transition always leads to a two-dimensional disordered polymer
adsorbed onto the surface, which only at higher temperature becomes three-dimensional.
This fact could be of particular relevance to polymer crystallization from solution. 
It suggests that a polymer forms a two-dimensional configuration on the growth surface before 
crystallizing, rather than crystallizing directly from solution. 
The thermodynamics of the crystallization process will depend significantly on which of these 
two possible mechanisms holds. 
The direct mechanism is implicitly or explicitly assumed in several theoretical descriptions.\cite{theorynoteb}
Yet, the present results appear to support the assumption made by Yamamoto in his study of 
the dynamics of the crystallization process.\cite{Yamamoto97} 

However, it is not clear whether this aspect of our results---crystallization being preceded 
by adsorption---is likely to be representative of the behaviour of real polymers; 
it might reflect some of the simplicities in the model. 
For example, in our model an adsorbed disordered polymer and a crystallite can have the 
same energy of interaction with the surface, whereas in reality a polymer chain that 
fits snugly into a groove formed by chains in the crystalline surface will have a lower 
energy than a random configuration on the surface. 
Such features could result in crystallization and adsorption occurring simultaneously.
Therefore, it would be useful if more realistic simulations were performed to 
clarify this situation.

In the phase diagram in Figure \ref{fig:phase} we have not attempted to distinguish 
regions where the scaling behaviour of the disordered state is that of a coil or of a globule. 
The main differences in behaviour between polymers with different values of $\epsilon_g$ 
are related to the position of the coil-globule transitions with respect to the adsorption
transition and show up most clearly in quantities such as the radius of gyration. 
For example, the maximum in $R_g^2$ for the two-dimensional  disordered polymer that occurs at 
$\epsilon_g=4\epsilon$ (Figure \ref{fig:rg}) does not occur for all $\epsilon_g$.
However, we do not pursue these issues further here.

\section{Free energy for a single crystallization pathway}
\label{sect:profiles}

Having understood the basic thermodynamics of our system we are now in a position to 
use the model to examine in detail the free energy landscape for the 
microscopic mechanism of polymer crystallization
assumed by the Lauritzen-Hoffman surface nucleation theory.\cite{Lauritzen60,Hoffman76a,Hoffman97} 
In this theory the key process
in determining the thickness of the lamellar crystals is considered to be the 
nucleation and growth of a new polymer layer on the crystal growth face (the thin
edges of the lamellae).
In the theory the growth rate for new layers of different thicknesses are compared.
It is argued that crystals with a thickness close to that which gives the 
maximum growth rate will dominate the ensemble of possible crystals, 
and so the predicted thickness of the crystal corresponds approximately to the 
thickness for which the rate is a maximum.

\begin{figure}
\begin{center}
\epsfig{figure=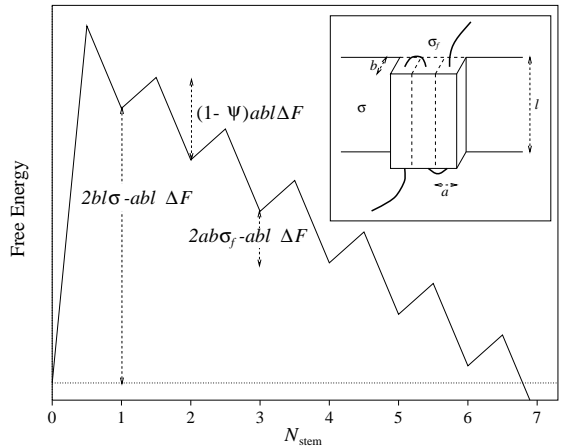,width=8.2cm}
\vglue0.1cm
\begin{minipage}{8.5cm}
\caption{\label{fig:snuc} Free energy profile assumed by the Lauritzen-Hoffman surface nucleation theory
of polymer crystallization. The theory considers the crystal to form by
laying down of adjacent stems of the polymer. The inset is a schematic
representation of a configuration where three stems have been deposited.
The profile is for a temperature at which the crystal is the most stable state}
\end{minipage}
\end{center}
\end{figure}

To calculate the growth rates some assumptions have to be made about the processes 
involved and the associated thermodynamics.
Firstly, the new layer is assumed to grow by the addition of a succession of stems
along the growth face each connected to the previous by a tight fold.
Secondly, the length of each stem is assumed to be the same as the thickness of 
the lamella.
Thirdly, to calculate the free energy for this process
macroscopic thermodynamic properties are used. 
This allows the free energy as a function of the number of stems to be written as 
\begin{equation}
A_L(N_{\rm stem})=2bl\sigma+ 2(N_{\rm stem}-1) a b \sigma_f - abl\Delta F,
\end{equation}
where $\sigma$ is the surface free energy of the growth surfaces, $\sigma_f$ is the
free energy of the fold surfaces, $\Delta  F$ is the free energy of crystallization
per unit volume, $l$ is the thickness of the lamella, $a$ is the width of a stem and
$b$ is the depth of a stem. The geometry of the assumed mechanism is depicted in the 
inset of Figure \ref{fig:snuc}. 
The first term in the above equation corresponds to the creation of the two new 
lateral surfaces on either side of the nucleus. This free energy has to be `paid for' on the 
laying down of the first stem. The second term corresponds to the free energy of the folds
that have to be created on the deposition of subsequent stems.

The fourth assumption is that at the barrier between states with different $N_{\rm stem}$ 
all the new surfaces have been created and that a fraction $\Psi$ of the free energy of 
crystallization has been released. These assumptions lead to free energy profiles like that
depicted in Figure \ref {fig:snuc}.

There have been a number of specific criticisms of this free energy profile.
Firstly, the model assumes that the value of $\Psi$ for deposition of the first 
and subsequent stems is the same when clearly the physical processes for these two cases are very different. 
The main motivation for this assumption is simply that it allows an analytical solution to the theory.
More recently, there has been some work which has attempted to find approximate
solutions when this simplifying assumption is relaxed.\cite{Snyder96a,Snyder97a}

Secondly, for any non-zero value of $\Psi$ the predicted thickness becomes infinite at
sufficiently large supercoolings---the so-called $\delta l$-catastrophe. 
This occurs when there is no longer any barrier for the deposition of the first 
stem; i.e.\ for temperatures where $\Psi>2\sigma/(a\Delta F)$. 
To avoid this unwanted behaviour, the theory has been supplemented with 
a justification for a zero value of $\Psi$. 
The argument is that the chain first forms a weakly physisorbed aligned state 
which traverses the growth face (such a configuration has a high free energy because it
has lost the entropy associated with orientational disorder)
before it crystallizes to form the first 
crystalline stem,\cite{Hoffman97} rather than the free energy of crystallization being
released as successive units of the first stem are deposited onto the surface.
There seems little evidence to favour the first scenario except that it avoids a
$\delta l$-catastrophe appearing in the theory.
Furthermore, the second mechanism would be expected to involve a lower free energy 
pathway, and so, if feasible, would be preferred.

If simulations are to help us make some critical assessment of the free energy profile used 
in the surface nucleation theory we need to devise an order parameter which can act as a reaction
coordinate along a pathway to a specific crystal. 
We use $N_{\rm xtal}$ which we define as the number of units in the largest fragment of the 
polymer that has part of the structure of our target crystal. 
This target crystal has adjacent re-entry of the stems which are are all of a specified length, $l$.
As our reaction coordinate is more fine-grained than that used in the surface nucleation 
theory it allows us to find the location of any free energy transition states on the path.

To compare with the theoretical free energy profile the pathway should go from the disordered state
of the polymer to the target crystal. However, the difficulty with the 
order parameter, $N_{\rm xtal}$, is that it is not able to distinguish between
configurations which are disordered and those that have a crystalline structure
different from the target crystal. 
Therefore, we have to introduce some constraints to ensure that the $N-N_{\rm xtal}$ units
which are not part of the largest fragment of the target crystal adopt a disordered configuration.
We use the following order parameters to achieve this:
$Q_{2D}^{\rm rest}$, $Q_{||}^{\rm rest}$ and $Q_{xr}$.
$Q_{2D}^{\rm rest}$ is simply the same as $Q_{2D}$ except that only bonds in the
non-crystalline part of the polymer are taken into account.
$Q_{||}^{\rm rest}=n_{||}^{\rm rest}/2 N_{\rm rest}$,
where $n_{||}^{\rm rest}$ is the number of instances when a bond in the 
non-crystalline part of the polymer is adjacent and parallel to another bond on the surface, 
and $N_{\rm rest}=N-N_{\rm xtal}$. 
$Q_{xr}=n_{xr}/N_{\rm xtal}$, where $n_{xr}$ is the number of contacts between the crystalline part 
of the chain and the rest of the polymer.

The constraints we use are: $Q_{2D}^{\rm rest}<0.3$, $Q_{||}^{\rm rest}<0.3$ and $Q_{xr}<0.15$.
Low values of $Q_{2D}^{\rm rest}$ and $Q_{||}^{\rm rest}$ are appropriate for a disordered chain. 
(The upper bounds have been found to be reasonable from monitoring these 
order parameters for the purely disordered state.)
The constraint on $Q_{xr}$ prevents the nucleus being stabilized by a more 
locally ordered region surrounding it.
However, the exact values for these constraints are slightly arbitrary. 
The free energy profiles that we obtain do reflect these choices to a certain extent, 
but only in the numerical details and not in their basic structure.

\begin{figure}
\begin{center}
\epsfig{figure=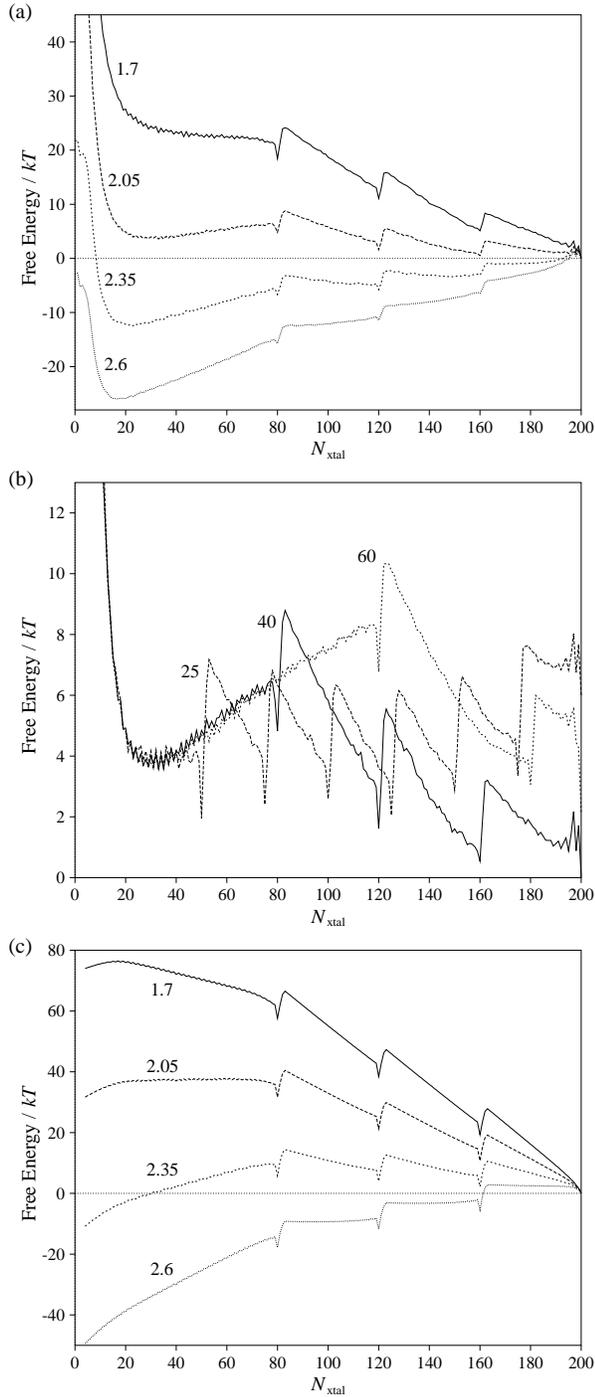,width=8.2cm}
\begin{minipage}{8.5cm}
\caption{\label{fig:free}Free energy for the formation of a target crystal with 
equal stem lengths and adjacent re-entry. 
In (a) and (c) the target crystal has stems 40 units long; 
the four curves correspond to different temperature, as labelled, 
and we have set the zero of free energy to that for the complete crystal.
In (b) the three curves correspond to target crystals with different stem length, 
as labelled, all at $T=2.05\epsilon k^{-1}$; 
we have set the zero of free energy to that for the complete crystal with stem length forty, and 
the free energy differences between the different complete crystals were calculated by 
complete enumeration of possible configurations.
The free energy profiles in (a) and (b) are from simulation and those
in (c) have been calculated using equation (\ref{eq:simple}).
}
\end{minipage}
\end{center}
\end{figure}

In Figure \ref{fig:free}a we show some free energy profiles that we obtain
for the formation of a target crystal with stems that are forty units long
at a variety of temperatures.
It is immediately apparent that the profiles have a sawtooth structure 
similar to the surface nucleation profiles. The minima occur at $2l$, $3l$ and
$4l$. Presumably, for longer polymers this structure would be repeated
at higher multiples of $l$. (Indeed when we use $l=25$ minima up to $7l$ are
observed (Figure \ref{fig:free}b).)
After the minimum the free energy rises sharply with the incorporation of the 
next two polymer units into the crystal reaching a maximum at $n l+2$ or $n l+3$.
These steps in the free energy have a clear interpretation. 
Configurations with a complete number of stems 
are favoured because such a configuration maximizes the number of polymer-polymer
contacts relative to the number of folds in the crystal. 
To increase the size of the crystal fragment a new fold has to be formed with 
an accompanying rise in the free energy. After this there is a monotonic decrease
in the free energy as the new stem grows until the next fold must form.
Translating this into the language used in the surface nucleation approach, 
$\Psi({N_{\rm stem}\rightarrow N_{\rm stem}+1})\approx 0$ for $N_{\rm stem}\ge 2$ and
the transition state occurs virtually immediately after the completion of the previous stem.

\begin{figure}
\begin{center}
\epsfig{figure=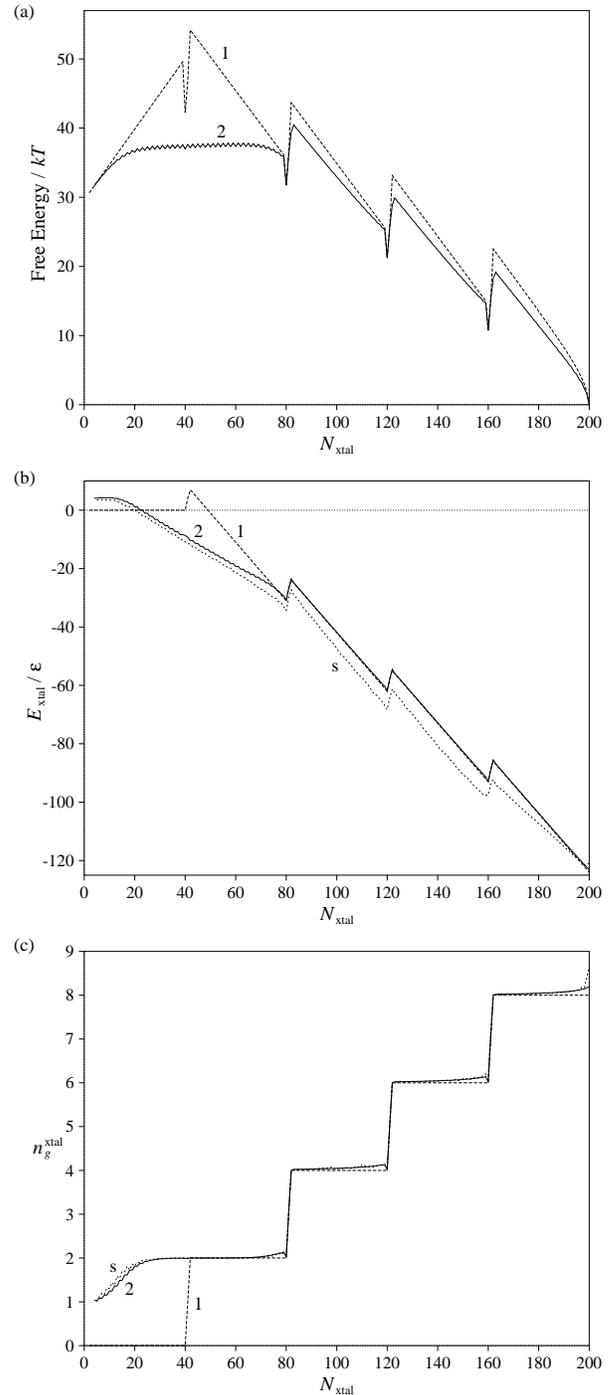,width=8.2cm}
\begin{minipage}{8.5cm}
\caption{\label{fig:theory.comp}
Comparison of results calculated using equation (\ref{eq:simple}) for crystallization pathways 
which allow one (dashed line 1) and two (solid line 2) incomplete stems to simulation (dotted line s) results.
(a) Free energy for the formation of the crystal,
(b) energy of the crystal (excluding the contribution from interactions
between the polymer and the surface), $E_{\rm xtal}$
(c) number of `gauche' bonds in the crystal, $n_g^{\rm xtal}$ as a function
of the number of units in the crystal.
The crystal has stems forty units long. $T=2.05\epsilon k^{-1}$
}
\end{minipage}
\end{center}
\end{figure}

However, at $N_{\rm xtal}<2l$ there are major differences between the simulation 
and the surface nucleation free energy profiles.
Firstly, there is no sign in the simulation results of any feature at $N_{\rm xtal}=l$ 
(or at any other value of $N_{\rm xtal}$) due to the formation of the first fold. 
Therefore, the mechanism of its formation must be significantly different from 
that for subsequent folds; it does not occur abruptly at a specific value of $N_{\rm xtal}$. 
In the surface nucleation approach it is assumed that new stems are laid down one at a time.
It does not allow for the possibility that two new stems could be formed simultaneously.
There is no such restriction in our simulations. 
We find that the initial nucleus is not a single stem but two incomplete stems of approximately equal length
connected by a fold. As $N_{\rm xtal}$ increases the two stems grow in length simultaneously.
The reason for this behaviour is energetic. 
It is simple to show that a two-stem nucleus can have a lower energy than a single-stem nucleus
when $N_{\rm xtal}>4\epsilon_g/\epsilon +2$. 
Confirmation of this behaviour can be found from Figure \ref{fig:theory.comp}c.
The number of gauche bonds in the crystal has risen to two by $N_{\rm xtal}\sim 25$.
The possibility of such a two-stem nucleus has previously been suggested by Point.\cite{Point79b}

We should note that it is possible that the size of the surface may have a small effect
on the competition between the two possible nuclei.
On our `infinite' surface the position of the nucleus is irrelevant.
However, on the growth surface of a lamellar crystal it is important that
the fold of the two-stem nucleus is close to (or at) the edge of the lamella,
whereas the initial position of a single-stem nucleus is unimportant.
Having said this a similar preference for two-stem nuclei has been observed 
in kinetic Monte Carlo simulations of polymer crystallization on the growth 
surface of a lamellar crystal.\cite{Doye98b}

Secondly, the behaviour at small $N_{\rm xtal}$ is different. 
For example at $T=1.7\epsilon k^{-1}$ the free energy is downhill until the barrier at $N_{\rm xtal}=2l$.
For the higher temperature free energy profiles there is a free energy minimum corresponding to the 
disordered state but it occurs at a finite value of $N_{\rm xtal}$, which decreases with increasing temperature. 
At $T=2.05, 2.35$ and $2.6\epsilon k^{-1}$ the minima are at $N_{\rm xtal}=33, 23$ and 17, respectively. 
Furthermore the free energy rises very steeply as $N_{\rm xtal}$ decreases towards zero. 
These effects occur because crystallization is occurring from a disordered state 
which is adsorbed onto the surface. 
This disordered state has structural motifs---single straight sections and 
short folds---which correspond to small fragments of the target crystal. 
For example the two-dimensional disordered conformation shown in Figure \ref{fig:piccies}b has
$N_{\rm xtal}=26$. 
As the temperature decreases the persistence length of the polymer
increases and so the value of $N_{\rm xtal}$ associated with the coil increases.

When there is a free energy minimum corresponding to the coil, 
for values of $N_{\rm xtal}$ beyond the minimum the free energy rises linearly until $N_{\rm xtal}=2l$.
Therefore, in this temperature range the barrier to forming a crystal nucleus 
with two complete stems does increase with $l$, as is clearly shown in Figure \ref{fig:free}b. 
However, this is very different from the $l$-dependence of the initial free energy barrier 
that is predicted by surface nucleation theory;
the top of the initial surface nucleation barrier occurs near to $N_{\rm stem}=1$ 
($N_{\rm xtal}=l$) and is a much steeper function of $l$.

To gain some further insight into the free energy profiles obtained from the simulation, 
we attempt to model them using a simple calculation in which we only explicitly consider
the crystalline part of the polymer configuration and assume the rest behaves like
an ideal two-dimensional coil. The free energy is then
\begin{equation}
\label{eq:simple}
A(N_{\rm xtal})=A_{\rm coil}(N_{\rm rest}) + \sum \exp \left( -\beta(E_{\rm xtal} + E_{\rm join}) \right),
\end{equation}
where the sum is over all possible configurations for the crystalline portion of the chain 
which are $N_{\rm xtal}$ units long,
$E_{\rm xtal}$, the energy of the crystal, is given by equation (\ref{eq:energy}),
$A_{\rm coil}$, the free energy of the ideal two-dimensional coil, is given by 
\begin{equation}
A_{\rm coil}(N_{\rm rest})=-N_{\rm rest} kT \log(1+2\exp(-\beta \epsilon_g)) -N_{\rm rest}\epsilon,
\end{equation}
and 
$E_{\rm join}$ is the energy of any gauche bonds that might by necessity occur at the junction between the 
crystalline and the coil parts of the chain. (The end of an incomplete stem in the crystalline 
portion of the chain must either occur at an end of the chain or be followed by a `gauche' bond.)

Some free energy profiles obtained from this approach are shown in Figure \ref{fig:free}c.
They have a remarkably similar structure to the simulation results,
although there are two main discrepancies.
Firstly, our expression for $A_{\rm coil}$ is an underestimate of the true free energy because it
neglects the energetic contribution from contacts between different parts of the coil. 
Therefore, equation (\ref{eq:simple}) overestimates the stability of the crystal 
compared to simulation results at the same temperature.
Secondly, we do not take into account the fact that structural patterns corresponding to small crystal nuclei
naturally occur in the disordered state, and so the calculated profiles do not have a large rise in 
free energy at small values of $N_{\rm xtal}$.

The useful feature of this simple calculation is that it allows us 
to understand some of the physical origins of the features in the free energy profile more easily. 
For example, $A(N_{\rm xtal}=nl)$ is always significantly lower than $A(N_{\rm xtal}=nl-1)$ ($n\ge 2$); 
this sharp dip in the free energy is because for $N_{\rm xtal}=nl$ neither end of the crystalline portion 
has to be terminated by a `gauche' bond, i.e. $E_{\rm join}=0$.

More interestingly, it can give greater insight into the effects of having a two-stem nucleus.
In Figure \ref{fig:theory.comp} we compare the results when all possible 
crystalline configurations are considered in the sum of equation (\ref{eq:simple}) and when only those in which 
there is a single incomplete stem are considered. 
In the latter case the initial nucleus must be a single stem, 
and so the results should be much closer to the SN profile. 
They are; the relevant free energy profile in Figure \ref{fig:theory.comp}a now shows a
free energy barrier for the formation of the first fold at $N_{\rm xtal}=l$ and
there is a steep initial rise in the free energy. 
However, it can be clearly seen that the free energy 
is much lower when the possibility of a two stem nucleus is allowed;
this is because the energy of a sufficiently-long two stem nucleus 
is lower than a single stem (Figure \ref{fig:theory.comp}b).
Furthermore, when assuming a single-stem nucleus the results for $E_{\rm xtal}$ and 
$n_g^{\rm xtal}$ (Figure \ref{fig:theory.comp}c),
as well as the free energy, significantly differ from simulation results, 
whereas the calculation and simulation are in good agreement when the two-stem nucleus is allowed.
(The slightly lower value of $E_{\rm xtal}$ for the simulation results is due to contacts between the crystalline
portion and the rest of the chain. These are not taken into account in our simple calculation.)

The calculated profiles have two slopes for $N_{\rm xtal}<2l$.
For small values of $N_{\rm xtal}$ the profile is steeper (and always positive)
because it is still more favourable to have a single-stem nucleus. (This feature
has no parallel in the simulation profile; instead, as discussed earlier there is a rapid rise in free energy 
as $N_{\rm xtal}$ becomes smaller.)
In this range the two calculated profiles in Figure \ref{fig:theory.comp}a are very similar.
In the range $N_{\rm xtal}=15-20$ the profile changes slope as the structure of the
nucleus changes.
Assuming an ideal two-stem nucleus the slope of the free energy profile for $20\lesssim N_{\rm xtal} < 2l$ 
will be
\begin{equation}
{dA\over d N_{\rm xtal}}=-{\epsilon\over 2}+kT \log(1+2 \exp(-\beta \epsilon_g)
\end{equation}
Therefore, the profile in this region is expected to be flat for 
$T=2.03\epsilon k^{-1}$ (Figure \ref{fig:theory.comp}a)
and below this temperature the calculated profile will exhibit a maximum at small values of $N_{\rm xtal}$ 
(e.g.\ $T=1.7\epsilon k^{-1}$ Figure \ref{fig:free}c).
A similar temperature-dependence of the slope for $N_{\rm xtal}\sim 20-80$ is seen in the simulation
free energy profiles (Figure \ref{fig:free}a).

\section{Conclusions}

The thermodynamic properties that we find for our semi-flexible polymer in the presence of a 
surface have many similarities to an isolated semi-flexible 
polymer,\cite{Kolinski86a,Doniach96a,Bastolla97a,Doye98a} e.g.\ the
coexistence of chain-folded crystallites with different aspect ratio 
and the order-disorder transition.
(It should be remembered that the formation of folded structures for our single polymer
system is a thermodynamic effect, whereas lamellar polymer crystals have folded chains due 
for kinetic reasons.)
One new feature is the desorption transition\cite{Eisenriegler93} in which the 
adsorbed two-dimensional polymer gradually adopts a three-dimensional configuration. 
Also of interest is the fact that crystallization is always preceded by adsorption.
However, it is not yet clear whether this feature is likely to be common to many polymers or 
reflects some of the aspects of our current model.

The free energy profiles that we obtain by umbrella sampling for specific crystallization 
pathways have a sawtooth structure where each rise in free energy corresponds to the
formation of a new fold. 
We found that the apportionment factor, 
$\Psi({N_{\rm stem}\rightarrow N_{\rm stem}+1})\approx 0$ for $N_{\rm stem}\ge 2$
($N_{\rm stem}$ is the number of stems in the crystalline configuration).
However for $N_{\rm stem} < 2$ our free energy profiles are significantly different from those
of Lauritzen and Hoffman's surface nucleation theory. 
There is no feature in our profiles that corresponds to the
formation of the first fold because the initial nucleus is not a single stem but two
incomplete stems connected by a single fold. 
The nucleus increases in size by the simultaneous growth of both stems. 
The reason that a two-stem nucleus is preferred is simply because beyond a certain size
it has a lower energy than a single-stem nucleus. 
Such energetic considerations are also likely to hold for more realistic polymer models.
This finding has serious implications for the coherence of the 
surface nucleation theory of polymer crystallization because the barrier to the formation of the
first stem, and its dependence on crystal thickness, plays a key role in the theory.

Our results also hint at another potential problem with the surface nucleation theory. 
To obtain the free energy profiles of Figure \ref{fig:free} we had to constrain the 
system to prevent crystallization occurring by other pathways. 
In surface nucleation theory only a small subset (those where the stem length is constant)
of the multitude of possible pathways are taken into account.
However, a rigorous treatment requires that the full web of possible paths is 
explored.\cite{Point79a,Point79c,DiMarzio82a} Recent kinetic Monte Carlo simulations
which do just this present a radically different picture of the mechanism by which 
the thickness of lamellar crystals is determined.\cite{Doye98b}

\acknowledgements
The work of the FOM Institute is part of the research program of
`Stichting Fundamenteel Onderzoek der Materie' (FOM)
and is supported by NWO (`Nederlandse Organisatie voor Wetenschappelijk Onderzoek'). 
JPKD acknowledges the financial support provided by the Computational Materials Science 
program of the NWO.
We thank James Polson for a critical reading of the manuscript.

\end{multicols}
\end{document}